\documentclass[oribibl]{llncs}%

\usepackage{graphicx}%
\usepackage{overpic}%
\usepackage{subfigure}%
\usepackage{verbatim}%

\begin{document}

\title{Computing Naturally in the Billiard Ball Model}

\author{Liang Zhang}

\institute{Center of Unconventional Computing and Department of Computer Science
\\University of the West of England, Bristol, United Kingdom\\
\email{Liang3.Zhang@uwe.ac.uk}}

\maketitle

\begin{abstract}
Fredkin's Billiard Ball Model (BBM) is considered one of the fundamental models of collision-based computing, and it is essentially based on elastic collisions of mobile billiard balls. Moreover, fixed mirrors or reflectors are brought into the model to deflect balls to complete the computation.
However, the use of fixed mirrors is ``physically unrealistic'' and makes the BBM not perfectly momentum conserving from a physical point of view, and it imposes an external architecture onto the computing substrate which is not consistent with the concept of ``architectureless'' in collision-based computing.
In our initial attempt to reduce mirrors in the BBM, we present a class of gates: the $m$-counting gate, and show that certain circuits can be realized with few mirrors using this gate. We envisage that our findings can be useful in future research of collision-based computing in novel chemical and optical computing substrates.

\end{abstract}

\section{Introduction}
\label{intro}

Collision-based computing~\cite{cbc_2003} is an unconventional paradigm of computation, where quanta of information are represented by compact patterns or localizations traveling in spatially-extended architectureless medium, such as billiard balls in an idealized friction-free environment~\cite{fredkin:toffoli}, gliders in cellular automata~\cite{conway, margolus, adamatzky_1995, spiral_rule}, discrete solitons in optical waveguide lattices~\cite{d_soliton} and wave-fragments in a sub-excitable Belousov--Zhabotinsky medium~\cite{wave_frag}. The information is encoded in binary as the presence or absence of localizations corresponds to logical 1 and 0. The computation is performed by mutual collisions of localizations. Trajectories of localizations approaching a collision site represent input values; and trajectories of localizations traveling away from a collision site represent output values. Pre-determined stationary wires and gates are not needed in collision-based computing. Anywhere within the medium space can be used as wires (trajectories of traveling localizations) and gates (collective collision sites of localizations).

Fredkin's Billiard Ball Model~\cite{fredkin:toffoli} is one of the fundamental models of collision-based computing and a model of universal physical computation, based on idealized elastic collisions of mobile billiard balls. Additionally, fixed mirrors are used to route signals. As one may find that mirrors are almost unavoidable when realizing complicated circuits in the BBM. However, there are several reasons we want to remove mirrors from the model. First, as Margolus~\cite{ssm} stated, fixed mirrors are ``physically unrealistic'', since in reality perfectly fixed mirrors ``must be infinitely massive'', otherwise a sightly shift in position of a single mirror could result in the model losing its ``digital nature''.
Second, from the collision-based computing point of view, mirrors stand for external architecture onto the computing medium. Although they are not stationary wires that signals propagate along and the virtual wires formed by traveling balls are still dynamical, mirrors are routing devices that control the moving directions of signals. Therefore the model is not entirely architectureless. 
More importantly, for anyone who intends to use the BBM as a blueprint to guide their research in collision-based computing using substrates in the real world, it may not be easy to look for an analogy to mirrors.

Margolus~\cite{ssm} once successfully replaced mirrors with constant particle streams together with the dual-rail logic to deflect signals
in a lattice gas version of his Soft Sphere Model (SSM, a model very similar to the BBM, with the difference that balls are compressible during collisions), in an attempt to achieve momentum conservation. Also, a concept of rest particle was introduced into the model to fix the problem of signals crossing without mirrors. Obviously, the dual-rail logic does not apply to the BBM, and the rest particle has no room in the BBM simply because the BBM cannot be simulated by a lattice gas model. Thus methods to remove mirrors in the SSM cannot be directed applied to the BBM.

In this paper, we present a class of gates: the $m$-counting gate. By using the gate as building blocks, we demonstrate that certain circuits can be realized in the BBM with few mirrors in place, which is one step closer to a physical realistic, architectureless, and momentum conserving BBM. The structure of the paper is as follows, we first review the BBM in Sect.~\ref{bbm}. Then we introduce the $m$-counting gate in Sect.~\ref{m_counting}, and demonstrate how we use such a gate to realize binary adders and parallel binary counters in Sect.~\ref{adders} and Sect.~\ref{counters}. As a conclusion, we discuss the difference between the computational logic of the $m$-counting gate and primitive gates in the BBM, and how our findings may benefit the future research in Sect.~\ref{discussions}.

\section{The Billiard Ball Model}
\label{bbm}

\begin{figure}[tbp]
	\centering
		\subfigure[]{\includegraphics[scale=.75]{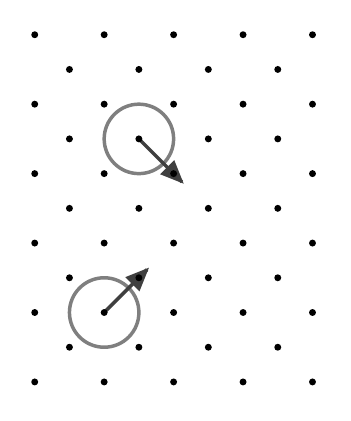}}
		\quad\subfigure[]{\includegraphics[scale=.75]{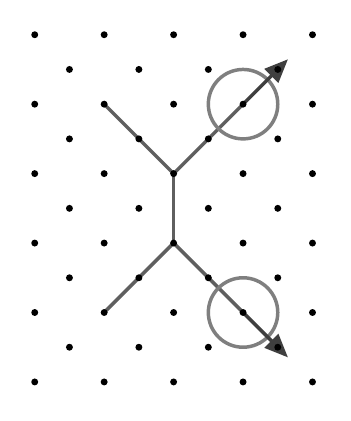}}
		\quad\subfigure[]{\includegraphics[scale=.75]{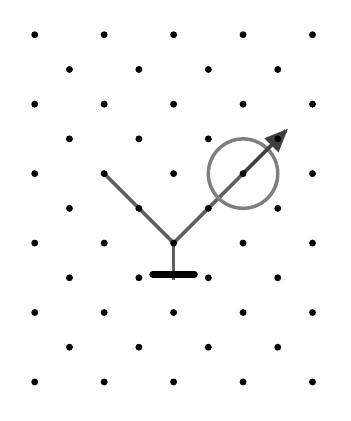}}
	\caption{The Billiard Ball Model.
	(a)~balls move diagonally on a two-dimensional grid;
	(b)~two balls collide and change their moving directions;
	(c)~a ball moving South-East is deflected by a fixed mirror and heads North-East instead.
In these graphs, arrows show moving directions of the balls, diagonal lines represent trajectories that balls traveled and vertical lines indicate collisions, either between two balls or between a ball and a mirror.}
	\label{BBM_intro}
\end{figure}

In the BBM, identical balls with finite diameter travel diagonally on a two-dimensional grid with the same magnitude of velocity. Balls are all initially placed on one of the dots of the grid, and the magnitude of their velocity is specifically chosen so that balls can appear at diagonally adjacent dots at every integer time (Fig.~\ref{BBM_intro}a). On two occasions balls can change their moving directions: colliding against each other (Fig.~\ref{BBM_intro}b) and/or being deflected by a fixed mirror (Fig.~\ref{BBM_intro}c). The diameter of balls is chosen to be the distance between two vertical (or horizontal) adjacent dots of the grid and the fixed mirrors should be carefully placed to ensure that collisions only happen when balls on one of the dots of the grid. Binary information, 1 and 0, is encoded by the presence or the absence of a ball at a given site of the grid at a given time. Therefore, trajectories of balls can be seen as routes of traveling signals, or wires. The routing of signals and the gates can be realized by collisions of balls (with the help of mirrors, sometimes).

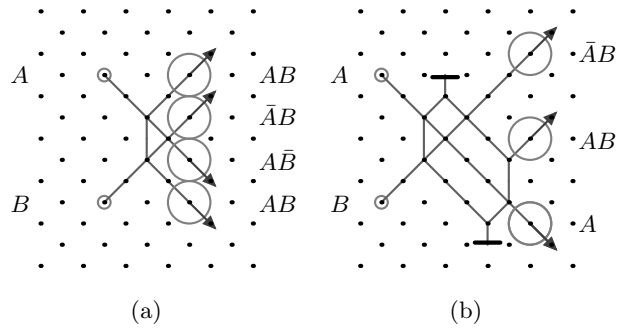
\begin{figure}[tbp]
\centering
\subfigure[]{
	\begin{overpic}[scale=.8]
		{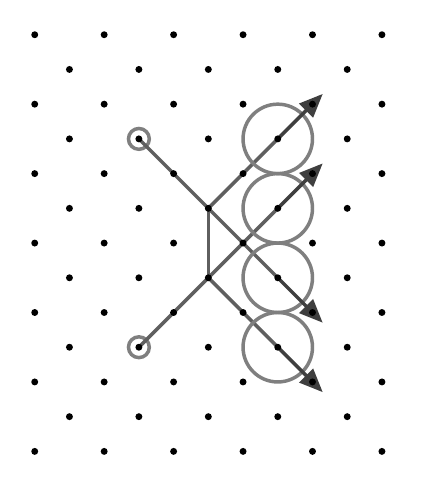}
		\put(-3,69){$A$}
		\put(-3,26){$B$}
		\put(80,69){$AB$}
		\put(80,55){$\bar{A}B$}
		\put(80,40){$A\bar{B}$}
		\put(80,26){$AB$}
	\end{overpic}
}
\qquad\subfigure[]{
	\begin{overpic}[scale=.8]
		{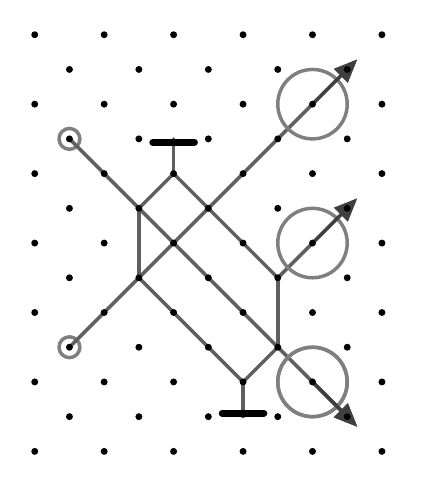}
		\put(-3,69){$A$}
		\put(-3,26){$B$}
		\put(80,76){$\bar{A}B$}
		\put(80,47){$AB$}
		\put(80,19){$A$}
	\end{overpic}
}
\caption{Primitive gates in the Billiard Ball Model: 
(a)~the interaction gate, and (b)~the switch gate.}
\label{primitive_gates}
\end{figure}

Two primitive gates were introduced in~\cite{fredkin:toffoli}, namely, the interaction gate and the switch gate (Fig.~\ref{primitive_gates}). 
The interaction gate is merely realized by a possible collision of two balls. Fig.~\ref{primitive_gates}a shows a superposition of all possible trajectories concerning the gate: when two balls start traveling from positions denoted by small circles on the left, they will collide with each other; and if only one ball is present at the starting positions, it will travel straight through.
The switch gate (Fig.~\ref{primitive_gates}b) is constructed on the basis of the interaction gate with two additional mirrors. The output signals $AB$ of the interaction gate are routed by mirrors so that one of them falls into the same route as the ball $A\bar{B}$, thus realizing the switch gate. It has been demonstrated that both gates are capable of universal computation.

Later works related to the BBM include a cellular automaton (BBMCA) \cite{margolus} that simulates the behavior described in the BBM and the Soft Sphere Model~\cite{ssm}, both developed by Margolus. Both models dealt with mirrors in a different approach to the BBM. In the BBMCA, mirrors are formed by the computing substrate itself, which is amazing because no external medium is involved. However, when experimenting on real media, it is still hard to find an analogy to the mirrors. In the SSM, especially in a lattice gas version of the model, Margolus managed to remove all of the mirrors. In the BBM, maybe we cannot remove them all, at least not until we find a way to overcome the problem of signals crossing. But we can try to reduce mirrors to a minimum level, and that is where the $m$-counting gate can help us in certain situations.

\section{The $m$-counting Gate}
\label{m_counting}

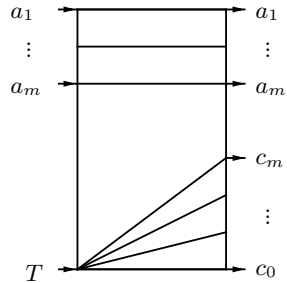
\begin{figure}[htbp]
\centering
	\begin{overpic}[scale=.7]
		{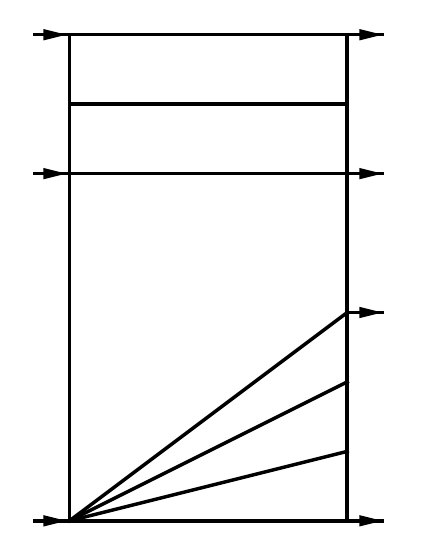}
		\put(-10,91){$a_{1}$}
		\put(-5,82){.}
		\put(-5,80){.}
		\put(-5,78){.}
		\put(-10,66){$a_{m}$}
		\put(72,91){$a_{1}$}
		\put(75,82){.}
		\put(75,80){.}
		\put(75,78){.}
		\put(72,66){$a_{m}$}
		\put(-5,3){$T$}
		\put(72,42){$c_{m}$}
		\put(75,26){$.$}
		\put(75,24){$.$}
		\put(75,22){$.$}
		\put(72,5){$c_{0}$}
	\end{overpic}
\caption{The graphical realization of the $m$-counting gate in the Billiard Ball Model, which can be used to count how many logical {\sc Truth}s there are in all $m$ input variables ($a_{1}$...$a_{m}$).}
\label{m-counting}
\end{figure}

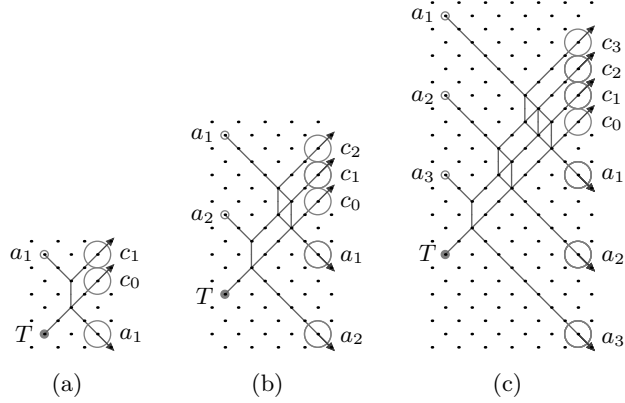
\begin{figure}[htbp]
\centering
\subfigure[]{
	\begin{overpic}[scale=.5]
		{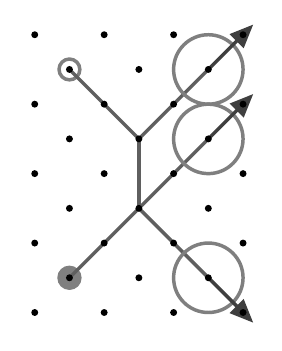}
		\put(-6,77){$a_{1}$}
		\put(-2,15){$T$}
		\put(76,77){$c_{1}$}
		\put(76,57){$c_{0}$}
		\put(76,17){$a_{1}$}
	\end{overpic}
}
\qquad\subfigure[]{
	\begin{overpic}[scale=.5]
		{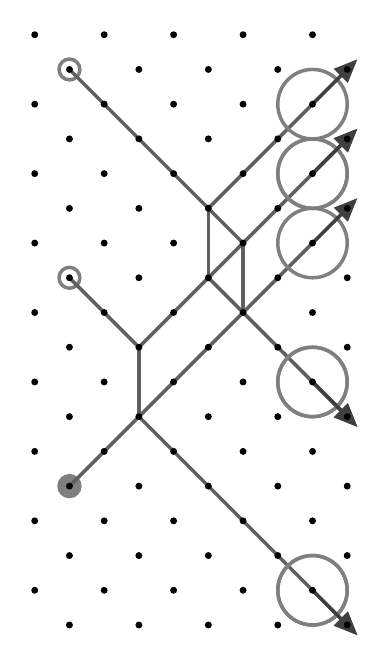}
		\put(-3,88){$a_{1}$}
		\put(-3,56){$a_{2}$}
		\put(-1,23){$T$}
		\put(56,82){$c_{2}$}
		\put(56,72){$c_{1}$}
		\put(56,61){$c_{0}$}
		\put(56,40){$a_{1}$}
		\put(56,8){$a_{2}$}
	\end{overpic}
}
\qquad\subfigure[]{
	\begin{overpic}[scale=.5]
		{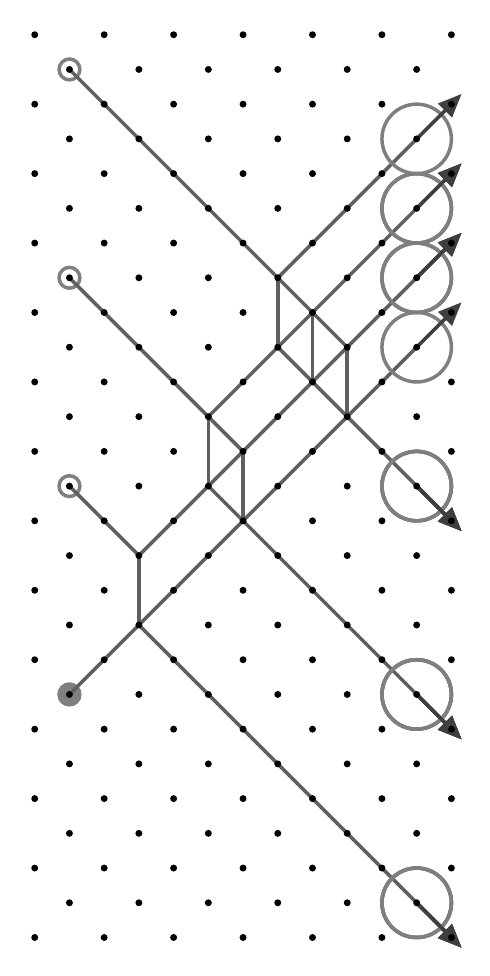}
		\put(-2,92){$a_{1}$}
		\put(-2,70){$a_{2}$}
		\put(-2,49){$a_{3}$}
		\put(0,27){$T$}
		\put(49,84){$c_{3}$}
		\put(49,77){$c_{2}$}
		\put(49,70){$c_{1}$}
		\put(49,63){$c_{0}$}
		\put(49,49){$a_{1}$}
		\put(49,27){$a_{2}$}
		\put(49,5){$a_{3}$}
	\end{overpic}
}
\caption{Simplest examples of the $m$-counting gate in the Billiard Ball Model: 
(a)~the 1-counting gate, where $c_{0} = \bar{a}_{1}$ and $c_{1} = a_{1}$
(b)~the 2-counting gate, where $c_{0} = \bar{a}_{1}\bar{a}_{2}$, $c_{1} = a_{1}\bar{a}_{2}+\bar{a}_{1}a_{2}$ and $c_{2} = a_{1}a_{2}$, and
(c)~the 3-counting gate, where $c_{0} = \bar{a}_{1}\bar{a}_{2}\bar{a}_{3}$, $c_{1} = a_{1}\bar{a}_{2}\bar{a}_{3}+\bar{a}_{1}a_{2}\bar{a}_{3}+\bar{a}_{1}\bar{a}_{2}a_{3}$, $c_{2} = a_{1}a_{2}\bar{a}_{3}+a_{1}\bar{a}_{2}a_{3}+\bar{a}_{1}a_{2}a_{3}$ and $c_{3} = a_{1}a_{2}a_{3}$. 
Initial positions of input variables ($a_{1}$...$a_{m}$) are represented by circles and that of the constant {\sc Truth} input $T$ is represented by a dot. Output variables include an exact copy of all input variables ($a_{1}$...$a_{m}$) and a single ball positioned at $c_{n}$, denoting that there are $n$ out of $m$ input variables being constant {\sc Truth}.}
\label{countings}
\end{figure}

The $m$-counting gate has $m$ input variables ($a_{1}$...$a_{m}$) and its function is to count the number of logical {\sc Truth}s $n$ out of all $m$ input variables. A graphical realization of the gate in the BBM is shown in Fig.~\ref{m-counting}, and realizations of three simplest instances of the gate in the BBM are shown in Fig.~\ref{countings}.

Generally, realization of the $m$-counting gate in the BBM consists of $m$ balls ($a_{1}$...$a_{m}$) as input variables moving in one direction, together with a ball $T$ representing a constant logical {\sc Truth} moving in another direction. 
The gate has {$2m+1$} output variables, with $m$ of which being an exact copy of all input variables ($a_{1}$...$a_{m}$) and other $m+1$ output variables ($c_{0}$...$c_{m}$). 
The traveling routes of the output variables are solely determined by the input balls, with $a_{1}$...$a_{m}$ moving one line to the South of their original traveling routes. 
The route of $c_{0}$ is always the same as the constant {\sc Truth} ball $T$, and routes of $c_{1}$...$c_{m}$ are $1$ to $m$ lines to the North of $c_{0}$. Among these $m+1$ output variables ($c_{0}$...$c_{m}$), only one of them, $c_{n}$, equals to logical {\sc Truth}, showing the result of the $m$-counting gate.
Depending on the number of input variable balls being logical {\sc Truth}: $n$, exactly $n$ collisions will happen during the computation, starting from the constant {\sc Truth} ball $T$ dynamically deflects the first ball it encounters. After each collision, the ball moving North-East changes its traveling route one line higher, as it would have done in a normal interaction gate~\ref{primitive_gates}a, so that after $n$ collisions, the output variable $c_{n} = 1$. If there is no ball to deflect at all, the ball $T$ remains its traveling route so that the output variable $c_{0} = 1$. 

\begin{figure}[htbp]
\centering
\subfigure[]{
	\begin{overpic}[scale=.5]
		{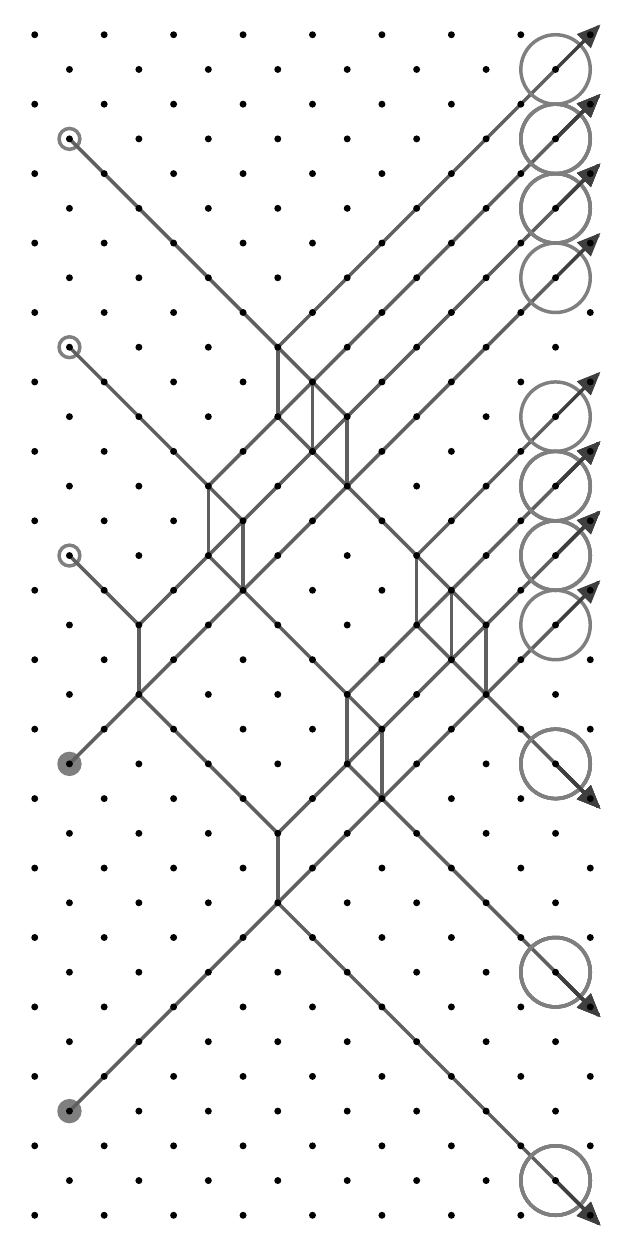}
		\put(-2,89){$a_{1}$}
		\put(-2,71){$a_{2}$}
		\put(-2,55){$a_{3}$}
		\put(-2,38){$T_{1}$}
		\put(-2,10){$T_{2}$}
		\put(49,94){$c_{3}$}
		\put(49,88){$c_{2}$}
		\put(49,83){$c_{1}$}
		\put(49,77){$c_{0}$}
		\put(49,66){$c_{3}$}
		\put(49,60){$c_{2}$}
		\put(49,55){$c_{1}$}
		\put(49,49){$c_{0}$}
		\put(49,38){$a_{1}$}
		\put(49,21){$a_{2}$}
		\put(49,5){$a_{3}$}
	\end{overpic}
}
\qquad\subfigure[]{
	\begin{overpic}[scale=.5]
		{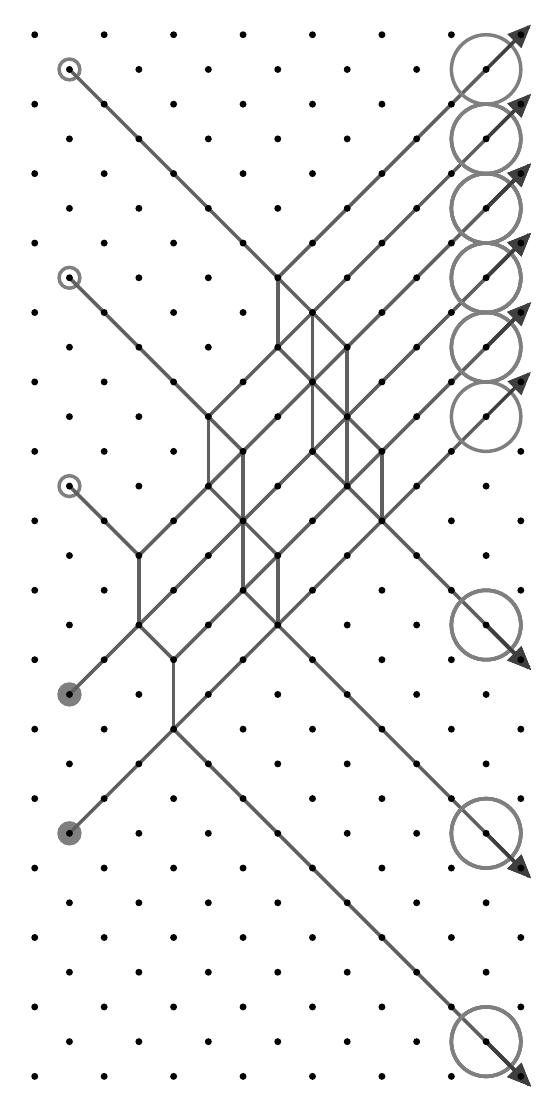}
		\put(-2,93){$a_{1}$}
		\put(-2,74){$a_{2}$}
		\put(-2,55){$a_{3}$}
		\put(-2,37){$T_{1}$}
		\put(-2,24){$T_{2}$}
		\put(49,94){$c_{3}$}
		\put(49,87){$c_{2}$}
		\put(49,80){$c_{3}+c_{1}$}
		\put(49,74){$c_{2}+c_{0}$}
		\put(49,68){$c_{1}$}
		\put(49,62){$c_{0}$}
		\put(49,43){$a_{1}$}
		\put(49,24){$a_{2}$}
		\put(49,5){$a_{3}$}
	\end{overpic}
}
\qquad\subfigure[]{
	\begin{overpic}[scale=.5]
		{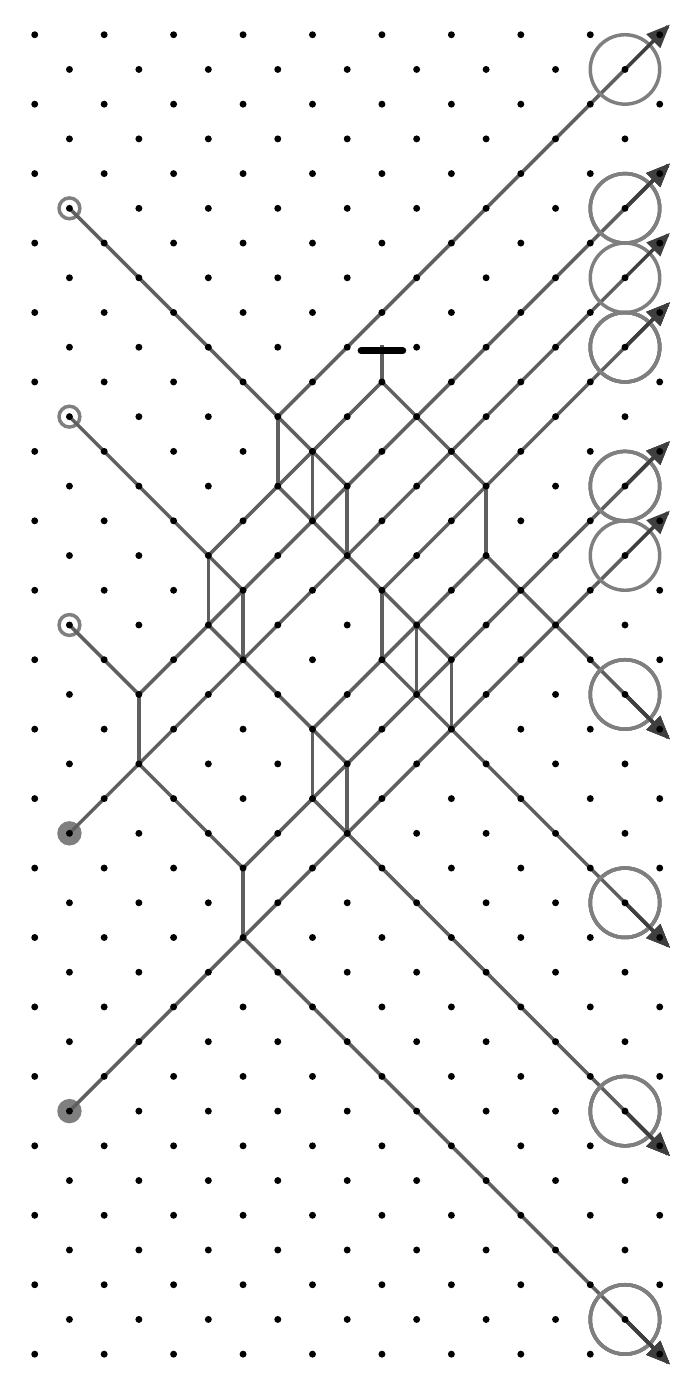}
		\put(-1,84){$a_{1}$}
		\put(-1,69){$a_{2}$}
		\put(-1,54){$a_{3}$}
		\put(-1,39){$T_{1}$}
		\put(-1,19){$T_{2}$}
		\put(49,94){$c_{3}$}
		\put(49,84){$c_{1}$}
		\put(49,79){$c_{0}$}
		\put(49,74){$c_{3}+c_{2}$}
		\put(49,65){$c_{1}$}
		\put(49,59){$c_{0}$}
		\put(49,49){$c_{2}$}
		\put(49,34){$a_{1}$}
		\put(49,19){$a_{2}$}
		\put(49,4){$a_{3}$}
	\end{overpic}
}
\caption{Multiple $m$-counting gate in the Billiard Ball Model:
(a)~two 3-counting gates generate two sets of outputs ($c_{0}$...$c_{3}$);
(b)~move the constant {\sc Truth} balls closer may generate outputs such as $c_{3}+c_{1}$, $c_{2}+c_{0}$;
(c)~however, to produce outputs in the form of $c_{n}+c_{n-1}$ needs the help of mirrors.}
\label{multiple}
\end{figure}
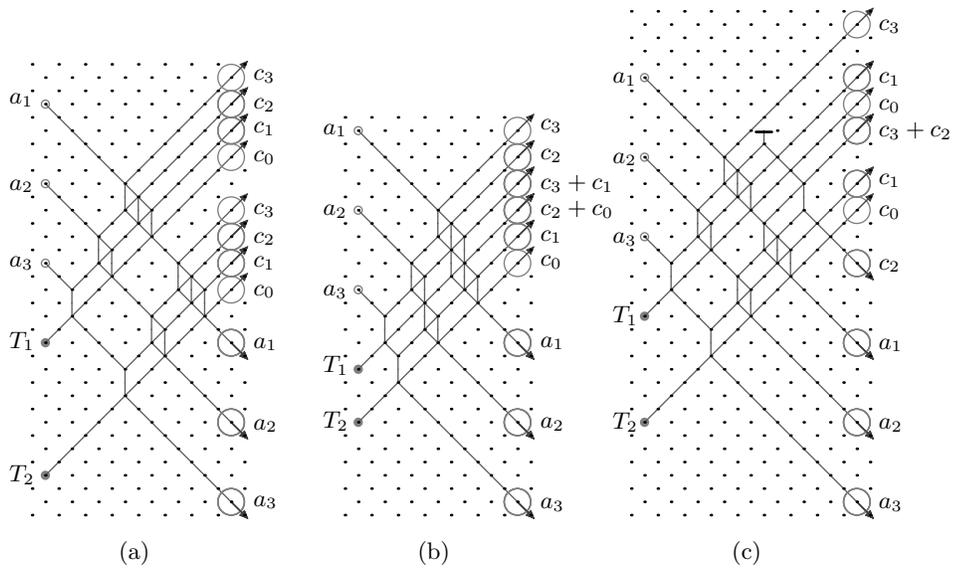

The feature of the $m$-counting gate that an exact copy of input variables is included in its output variables allows us to generate multiple instances of output variables ($c_{0}$...$c_{m}$) with only one set of input variables ($a_{1}$...$a_{3}$) and multiple instances of the constant {\sc Truth} balls.
As illustrated in Fig.~\ref{multiple}a, the input variables ($a_{1}$...$a_{3}$) interact with the first constant {\sc Truth} ball $T_{1}$ and generate a set of output variables ($c_{0}$...$c_{3}$), together with a copy of the input variables ($a_{1}$...$a_{3}$), which in turn interact with the second constant {\sc Truth} ball $T_{2}$, generating a second set of ($c_{0}$...$c_{3}$) as well as a copy of ($a_{1}$...$a_{3}$) again.

Relative positions of the multiple instances of output variables ($c_{0}$...$c_{m}$) are determined by relative positions of the constant {\sc Truth} balls. Thus by adjusting relative positions of adjacent constant {\sc Truth} balls, we can perform logical disjunction operation on certain output variables ($c_{0}$...$c_{m}$). As shown in Fig.~\ref{multiple}b, while $T_{1}$ and $T_{2}$ move closer, traveling routes of the two sets of output variables ($c_{0}$...$c_{3}$) overlap, resulting in two outputs $c_{3}+c_{1}$ and $c_{2}+c_{0}$.

Since constant {\sc Truth} balls cannot be too close to be next to each other, we cannot generate outputs in the form of $c_{n}+c_{n-1}$ directly. To do that, we need to use a mirror, as illustrated in Fig.~\ref{multiple}c. Again, two constant {\sc Truth} balls are used, producing two set of output variables ($c_{0}$...$c_{3}$). The mirror is placed to deflect one of the balls representing $c_{2}$ (the top one), which then collide with the other ball representing $c_{2}$ (the bottom one) and its traveling route end up the same as a ball representing $c_{3}$, resulting in an output $c_{3}+c_{2}$.

\section{Binary Adders}
\label{adders}

The $m$-counting gate is very helpful in realizing binary adders with few mirrors or even no mirrors at all in the BBM. The simplest binary adder, the 1-bit half adder, resides in the 2-counting gate (Fig.~\ref{countings}b), with two input variables $a_{1}$ and $a_{2}$ being summands and the output variables $c_{1}$ and $c_{2}$ equal the {\sc Sum} and {\sc Carry} value of the 1-bit half adder respectively. This circuit is constructed with no mirrors. Imagine if we were to build the same circuit using only interaction gates, there would have been no way to generate the {\sc Sum} value without using fixed mirrors.

\begin{figure}[tbp]
\centering
	\begin{overpic}[scale=.5]
		{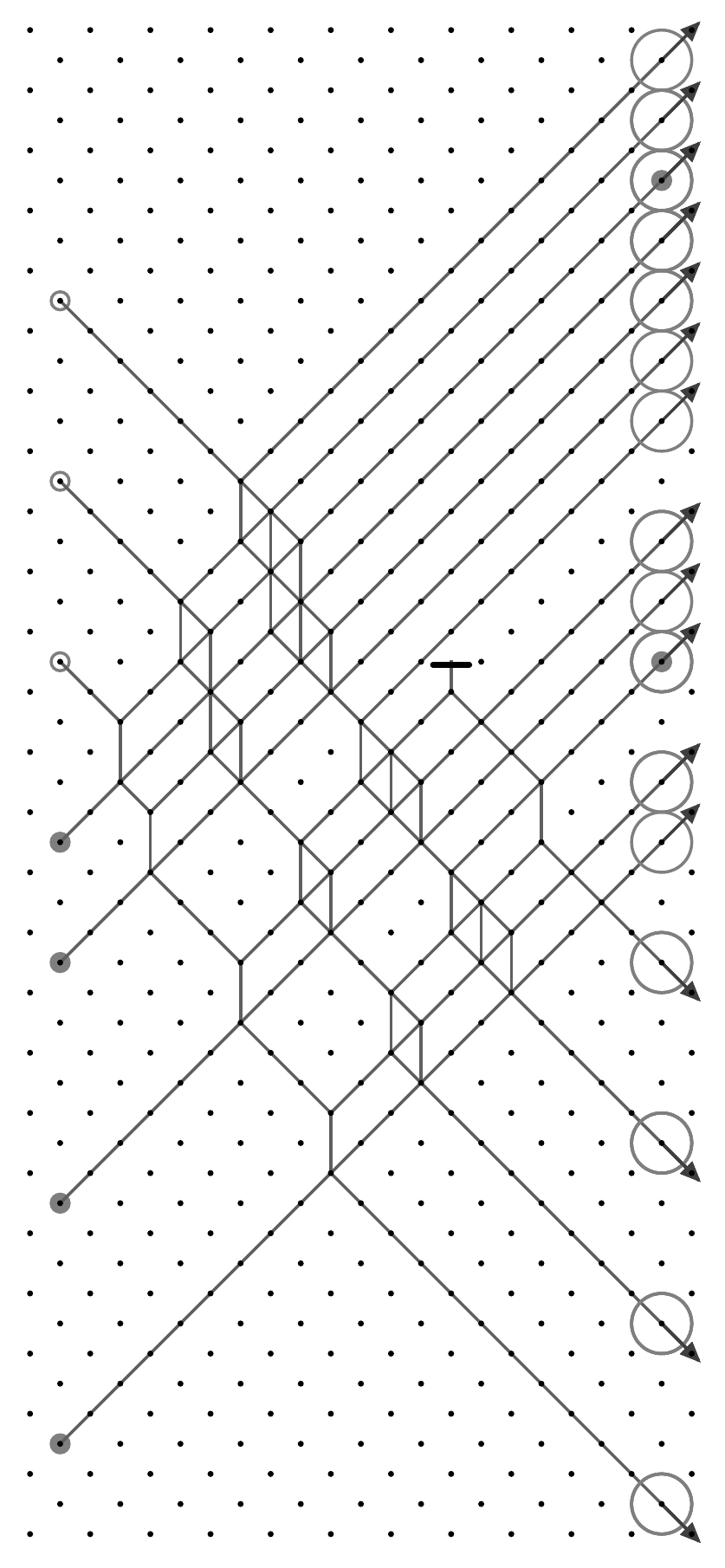}
		\put(-2,80){$a_{1}$}
		\put(-2,69){$a_{2}$}
		\put(-2,57){$a_{3}$}
		\put(-1,45){$T$}
		\put(-1,38){$T$}
		\put(-1,22){$T$}
		\put(-1,7){$T$}
		\put(45,87){$S = c_{3}+c_{1}$}		
		\put(45,56){$C_{out} = c_{3}+c_{2}$}
		\put(45,26){$a_{1}$}
		\put(45,15){$a_{2}$}
		\put(45,3){$a_{3}$}
	\end{overpic}
\caption{The realization of a binary 1-bit full adder using four 3-counting gates 
and a fixed mirror in the Billiard Ball Model}
\label{1-bit adders}
\end{figure}
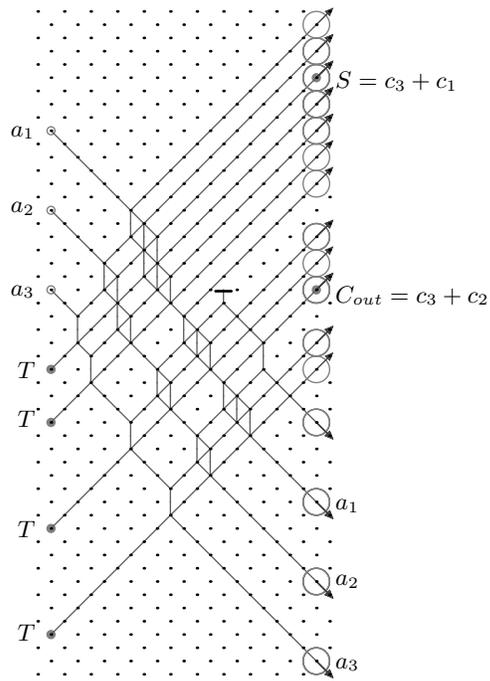

The binary 1-bit full adder, can be realized by four 3-counting gates and one extra mirror in the BBM, as illustrated in Fig.~\ref{1-bit adders}. It is actually a combination of the circuits in Fig.~\ref{multiple}b and ~\ref{multiple}c, since if we consider input variables ($a_{1}$...$a_{3}$) as two summands and the {\sc Carry-in} value of the adder, then the {\sc Sum} value $S = c_{3}+c_{1}$ and the {\sc Carry-out} value $C_{out} = c_{3}+c_{2}$.

\section{Parallel Binary ($m$,$k$)-counters}
\label{counters}

The concept of parallel binary ($m$,$k$)-counters was first introduced by Dadda~\cite{dadda65} to construct parallel multipliers in 1965. An ($m$,$k$)-counter has $m$ input variables, and it counts the number of input variables being logical {\sc Truth}s: $n$. This concept is very similar to the above $m$-counting gates, with the difference that in ($m$,$k$)-counters the result $n$ is recorded using $k$-bit binary numbers.

Three of the simplest instances of ($m$,$k$)-counters: (1,1)-counter, (2,2)-counter and (3,2)-counter are already realized in previous sections, since they are equivalent to the 1-counting gate (Fig.~\ref{countings}a), the 2-counting gate (Fig.~\ref{countings}b) and the 1-bit full adder (Fig.~\ref{1-bit adders}). 

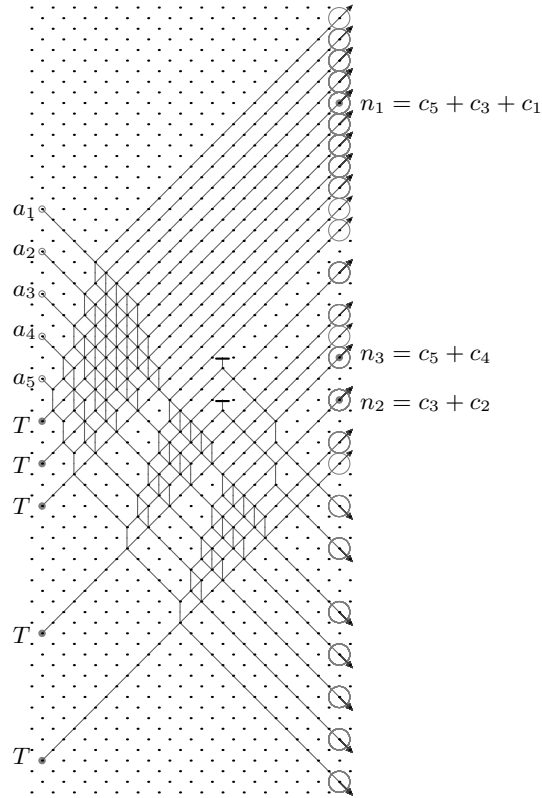
\begin{figure}[htbp]
\centering
	\begin{overpic}[scale=.4]
		{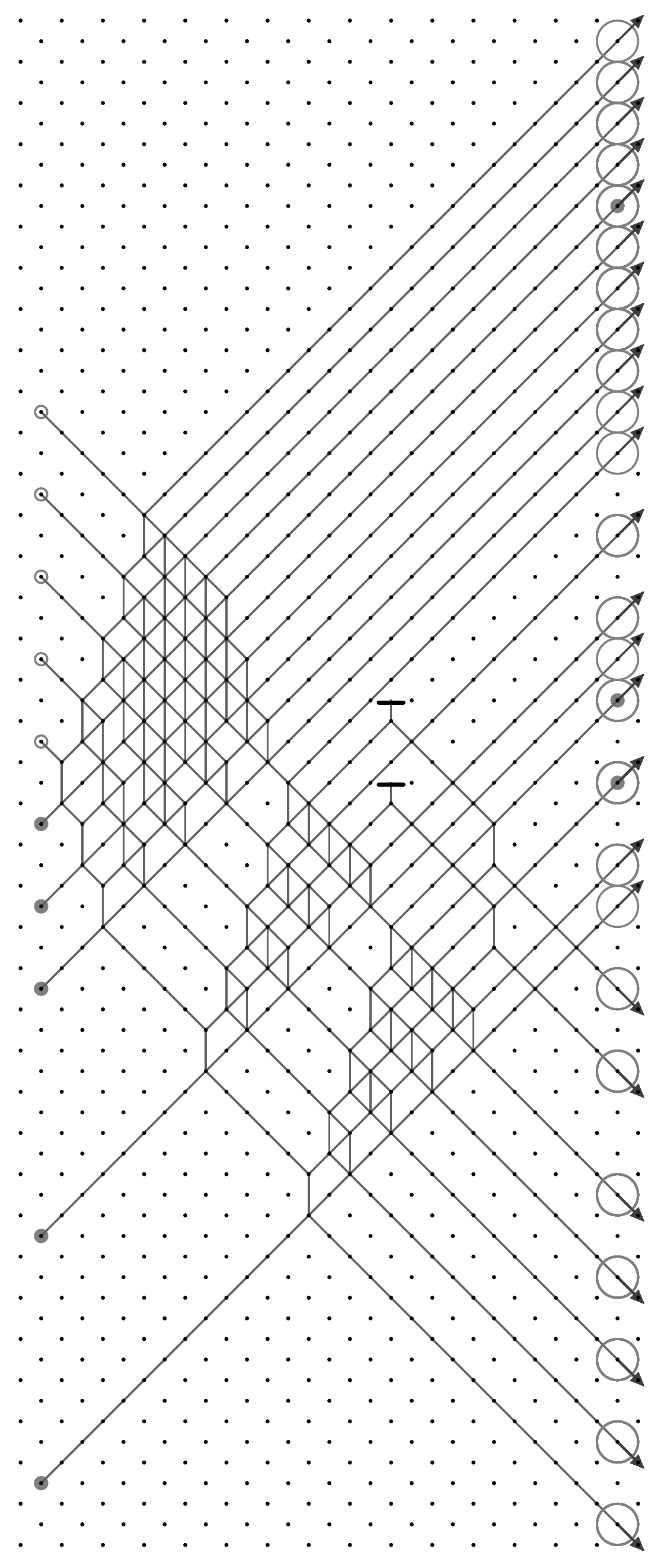}
		\put(-1,73){$a_{1}$}
		\put(-1,68){$a_{2}$}
		\put(-1,63){$a_{3}$}
		\put(-1,58){$a_{4}$}
		\put(-1,52){$a_{5}$}
		\put(-1,46){$T$}
		\put(-1,41){$T$}
		\put(-1,36){$T$}
		\put(-1,20){$T$}
		\put(-1,5){$T$}
		\put(42,86){$n_{1} = c_{5}+c_{3}+c_{1}$}
		\put(42,55){$n_{3} = c_{5}+c_{4}$}
		\put(42,49){$n_{2} = c_{3}+c_{2}$}
	\end{overpic}
\caption{A realization of the (5,3)-counter in the Billiard Ball Model, which uses five 5-counting gates and two fixed mirrors.}
\label{5_3_counter}
\end{figure}

Here we demonstrate a realization of the (5,3)-counter (Fig.~\ref{5_3_counter}), where the counting result $n$ is recorded in a 3-bit binary number: $n_{3}n_{2}n_{1}$.
The realization uses the same techniques that are used in the 1-bit full adder. First, three 5-counting gates are used to producing an output of $n_{1} = c_{1}+c_{3}+c_{5}$ by arranging three constant {\sc Truth} balls in close distances. Then, two sets of output variables ($c_{0}$...$c_{5}$) are generated by another two 5-counting gates, followed by the pairs of $c_{4}$s and $c_{2}$s colliding between themselves, producing the outputs of $n_{2} = c_{2}+c_{3}$ and $n_{3} = c_{4}+c_{5}$.

Other ($m$,$k$)-counters ($m \geq 6$) can also use $m$-counting gates as building blocks, however, more mirrors may be placed to route signals in order to avoid unwanted collisions.

\section{Discussions}
\label{discussions}
 
The $m$-counting gate reveals one of the underlying logic naturally existing in the BBM. Thus we can build certain circuits without recourse to lots of fixed mirrors, which is not possible when using only primitive gates --- the interaction gate and the switch gate. Although primitive gates are capable of universal computation, they do not naturally support disjunction operation or conjunction operation on more than two operands, which is why we often need multiple instances of such gates to realize a circuit and why we need a great many fixed mirrors to route signals. The switch gate itself is actually a good example of using mirrors to
perform a disjunction operation.

On the contrary, the $m$-counting gate produces outputs with several conjunction and disjunction operations on the inputs, avoiding those mirrors that may have been needed if we use primitive gates. Take the realization of parallel binary counters, for example, 
as described by Swartzlander~\cite{swartz}, conventionally an ($m$,$k$)-counter is implemented using a two-level gate network, which consists of $2^{m}$-1 {\small AND} gates with $m$ inputs as well as k {\small OR} gates with $2^{m}$-1 inputs. The $m$-counting gate do not build circuits from the simplest {\small AND} gates and {\small OR} gates, rather, some of those two-level gate networks are outputs of the $m$-counting gate.

The $m$-counting gate has its limitation. First, though it is suitable as the building blocks to construct binary adders and parallel binary counters, it may not be that efficient when used to build other circuits. Second, the $m$-counting gate produce many outputs, and not all of them may be useful in constructing other circuits. Thus those signals that are no use can turn into obstacles and need to be clear out of the space, which may need more fixed mirrors. Nonetheless, we present a new way of exploit the Billiard Ball Model and the $m$-counting gate shows us some insights into how the BBM can compute in its natural fashion without mirrors.

We envisage that our findings can be useful in future research of collision-based computing since the BBM is used as a blueprint of computing schemes in novel computing substrates, for example, the light-sensitive sub-excitable Belousov--Zhabotinsky (BZ) medium. Adamatzky, de Lacy Costello and their colleagues have demonstrated both computationally and experimentally that mutual collisions of wave-fragments in the sub-excitable BZ medium can implement certain logical gates and basic operations~\cite{bz_2004, bz_2005, bz_2007, bz_2008}. Further research in constructing complex circuits may require the implementation of ``mirrors'' by adjusting light intensity to control the traveling directions of wave-fragments, while our findings in the present paper show another approach to continue the research.

\section{Acknowledgment}
The author wishes to thank Andrew Adamatzky, Norman Margolus, Silvio Capobianco and Katsunobu Imai for their valuable comments and suggestions on this work.

\end{document}